\documentclass[aip,superscriptaddress,floatfix,reprint]{revtex4-1}
\usepackage{times}
\usepackage{graphicx}
\usepackage{amssymb}
\usepackage{amsmath}

\begin{document}
\title{TEM turbulence optimisation in stellarators}
\author{J.~H.~E.~Proll} 
\affiliation{Max Planck/Princeton Center for Plasma Physics}
\affiliation{Max Planck Institute for Plasma Physics, Wendelsteinstr. 1, 17491 Greifswald, Germany} 
\email{jproll@ipp.mpg.de}
\author{H.~E.~Mynick}
\affiliation{Plasma Physics Laboratory, Princeton University, P.O. Box 451 Princeton, New Jersey 08543-0451}
\author{P.~Xanthopoulos}
\affiliation{Max Planck Institute for Plasma Physics, Wendelsteinstr. 1, 17491 Greifswald, Germany}
\author{S.~A.~Lazerson}
\affiliation{Plasma Physics Laboratory, Princeton University, P.O. Box 451 Princeton, New Jersey 08543-0451}
\author{B.~J.~Faber}
\affiliation{HSX Plasma Lab, University of Wisconsin-Madison, Madison, WI 53706, USA} 
\affiliation{Department of Physics, University of Wisconsin-Madison, Madison, WI 53706, USA}

\date{\today}

\begin{abstract}
With the advent of neoclassically optimised stellarators, optimising stellarators for turbulent transport is an important next step. 
The reduction of ion-temperature-gradient-driven turbulence has been achieved via shaping of the magnetic field,
and the reduction of trapped-electron mode (TEM) turbulence is adressed in the present paper. 
Recent analytical and numerical findings suggest TEMs are stabilised when a large fraction of trapped particles experiences favourable bounce-averaged curvature.
This is the case for example in Wendelstein 7-X [C.D. Beidler {\it et al} Fusion Technology {\bf 17}, 148 (1990)] and other Helias-type stellarators.
Using this knowledge, a proxy function was designed to estimate the TEM dynamics, allowing
optimal configurations for TEM stability to be determined with
the STELLOPT [D.A. Spong {\it et al} Nucl. Fusion {\bf 41}, 711 (2001)] code without extensive turbulence simulations. A first proof-of-principle
optimised equilibrium stemming from the TEM-dominated stellarator experiment HSX [F.S.B. Anderson{\it et al}, Fusion Technol. {\bf 27}, 273 (1995)] is
presented for which a reduction of the linear growth rates is achieved over a broad
range of the operational parameter space. As an important consequence of this property, the turbulent heat flux levels are reduced compared with the initial configuration.
\end{abstract}

\maketitle
\normalsize
\section{Introduction}
The development of stellarators has taken great strides since their inception by Lyman Spitzer Jr. in 1951 \cite{Spitzer1951}. Stellarators are inherently 3-dimensional, so that the configuration space of possible equilibria is very large. 
Different techniques of optimisation have been employed to find equlibria with certain desired features within this configuration space. The reduction of neoclassical transport down to levels of tokamaks for example has been achieved through the introduction of 
quasi-symmetries \cite{Nuhrenberg1988,Boozer} or variants of omnigeneity \cite{Hall1975,Mynick1982}, which made stellarators competitive with tokamaks regarding the expected levels of transport. 
In addition, other stellarator design features are being addressed via optimisation, too, such as the confinement of fast particles \cite{Drevlak2014}.
In neoclassically optimised stellarators such as Wendelstein 7-X (W7-X)\cite{Beidler1990,Klinger2013} or the quasi-symmetric stellarators HSX (Helically Symmetric Experiment)\cite{Anderson1995} and NCSX (National Compact Stellarator Experiment)\cite{Zarnstorff2001}, the turbulent transport is expected to be the dominant transport channel in a large part of the plasma, just as in tokamaks. 
While NCSX has not been built and W7-X is only about to start operation, HSX is already running and has shown that the neoclassical transport is indeed reduced thanks to the quasi-symmetry \cite{Canik2007} and the turbulent transport gains importance.
This turbulent transport is thought to be driven by microinstabilities like the ion temperature gradient mode (ITG) or the trapped-electron mode (TEM). An optimisation to reduce these kinds of microinstabilities in stellarators is not trivial, especially since analytical predictions regarding the actual nonlinear behaviour of microturbulence in general geometry are rather sparse. 
For instance, only very recently Plunk \emph{et al.} \cite{Plunk2015} published a theory on the saturation of ITG turbulence whereas an analogous theory for TEMs has yet to be achieved.
Comprehensive numerical simulations of microturbulence in general geometry on the other hand have become available, although they are computationally very demanding - a typical well-resolved turbulence simulation in flux-tube geometry and including kinetic electrons needs roughly half a million CPUh to reach a saturated state. For this matter, an optimisation based on calculating the nonlinear heat flux as a figure of merit for every configuration along the path of optimisation would be desirable, 
but it is evident that this procedure is currently not viable. We must therefore find simplified expressions to represent the nonlinear heat flux as a figure of merit. Ideally, these ``proxies'' would be based on analytical theory of the linear or even nonlinear instabilities.
The reduction of ITG turbulence has been theoretically demonstrated via this method \cite{Mynick2010, Mynick2011, Mynick2014, Xanthopoulos2014} and in this paper we will tackle the reduction of TEM turbulence. Ultimately, one would want to combine all methods of optimisation 
into one grand scheme to find the point in configuration space where ``the ideal stellarator'' lives. A code that would be capable of carrying out such an ambitious task is STELLOPT \cite{Spong2002}, which we will also use for our TEM optimisation.
In the next section we briefly explain how STELLOPT works and how a new optimisation is implemented. In section \ref{sec:geometries} we review what we already know about TEMs in general geometry using analytical and linear numerical findings.
Section \ref{sec:proxies} then shows how we can use this knowledge to come up with simple expressions for the proxy. In section \ref{sec:HSXopt} we present a proof-of-principle equilibrium where we optimised starting from HSX towards reduced TEM turbulence and we comment on the applicability to
experimentally feasible TEM-optimised equilibria. Section \ref{sec:conclusions} contains the main conclusions, together with future plans related to this work.
\section{Optimising with STELLOPT}
\label{sec:STELLOPT}
The STELLOPT code is designed to optimise 3D MHD equilibria created by VMEC \cite{Hirshman1986} by minimising the difference between certain features of the equilibrium and their targeted value. Each design feature $i$ - this can be the neoclassical transport, the turbulent transport, ballooning stability, the major radius, to name only a few - 
is associated with a target value $f_{i}^{\rm target}$, and the difference between this target value and the actual value of the final ``optimised'' equilibrium $f_{i}^{\rm equilibrium}$ should be as small as possible. Since usually more than only one design feature shall be targeted at once,  each design feature gets assigned a tolerance $\sigma_i $. This acts as a weight ($1/\sigma^2$) when all design features are eventually combined into one function $\chi^2$ that has to be minimised:
\begin{equation}
\chi^2=\sum_i\frac{|f_{i}^{\rm equilibrium} -f_{i}^{\rm target}|^2 }{\sigma_i ^2}.
\end{equation}
How $f_i^{\rm equilibrium}$ for a given design feature in a given equilibrium is determined depends very much on the design feature itself. For the neoclassical transport for example, STELLOPT is coupled to the NEO code \cite{Nemov1999} to calculate the neoclassical
effective ripple $\epsilon_{\rm eff}$ \cite{Beidler2001}. For the turbulent transport it would be ideal to use the turbulent heat fluxes stemming from gyrokinetic simulations. However, since these are CPU-intensive, simpler proxy functions that can substitute for the turbulent heat flux
and that are ideally based on analytical theory are sought. 
The STELLOPT code can be used to optimise any combination of VMEC input parameters to any set of target figures of merit (FOM), subject to the constraints of a given optimisation method.  
When utilised for stellarator design the boundary harmonics are treated as the free parameters, although enclosed toroidal flux, net toroidal current and a pressure scaling factor may be included as well.  
As the VMEC boundary representation (R and Z harmonics) is non-unique, the initial configuration is converted to either Hirshman-Breslau \cite{Hirshman:1998ec} or Garabedian \cite{Bauer:1981tv} representation .  
These harmonics are the quantities varied by STELLOPT and converted back to the VMEC representation for evaluation of the configuration by VMEC.  
Once an optimum shape has been computed, codes like NESCOIL \cite{Merkel:1987vb} and COILOPT \cite{Strickler:2002wz,Zheng:2014eu} are utilised to generate a coil set consistent with that equilibrium.
To explore the space of accessible configuration in a given device, VMEC may be run in free boundary mode and STELLOPT set to treat the vacuum coil currents as free parameters.
Such a capability allows exploration of a given device's capabilities, as was done for NCSX \cite{Pomphrey:2007wx}.  
Additionally, the inclusion of plasma profiles, synthetic MSE diagnostics, and magnetic diagnostics \cite{Lazerson:2013dk} has allowed the code to provide a 3D equilibrium reconstruction capability \cite{Lazerson:2015bb,Schmitt:2014fy}.
In this paper, a modified Levenberg-Marquardt \cite{Marquardt:1963wr} algorithm was used to find the minimum of $\chi^2$ in configuration space. This method guarantees that the found optimised equilibrium is at least at a local minimum in configuration space. STELLOPT is also equipped with stochastic algorithms (e.g. differential evolution \cite{Goldberg1989}, particle swarm \cite{Kennedy:1995bi}) that have not been applied to this work.

\section{The density-gradient-driven TEM in general geometry}
\label{sec:geometries}
To tackle the optimisation of stellarators towards reduced TEM turbulence we should first assess what we already know about the TEM in general geometry. The TEM \cite{Kadomtsev1967,Dannert2005} can be regarded as a drift wave that is driven unstable by a resonance with the
precessional drift of trapped particles. Generally, the higher the trapped-particle fraction in a given configuration, the more unstable the TEM becomes. It is destabilised by increasing the density gradient and/or the electron temperature gradient. In a collisional plasma, trapped particles can become detrapped due to collisions, which usually leads to a stabilisation of the TEM \cite{Romanelli2007}.
Since this work addresses the suppression of the worst-case instability, collisions will be neglected from here on. First, we revisit some analytical theory regarding the stability properties of TEMs. Because of the temporal and spatial scales involved, the gyrokinetic framework is employed. We then look at different stellarator equilibria with very different geometric properties and present linear simulation results that confirm the analytical findings. 
More details on the calculations can be found in previously published papers \cite{Proll2012,Helander2012,Helander2013,Proll2013}.

\subsection{Analytical theory}
\begin{figure}
\includegraphics[width=0.5 \textwidth]{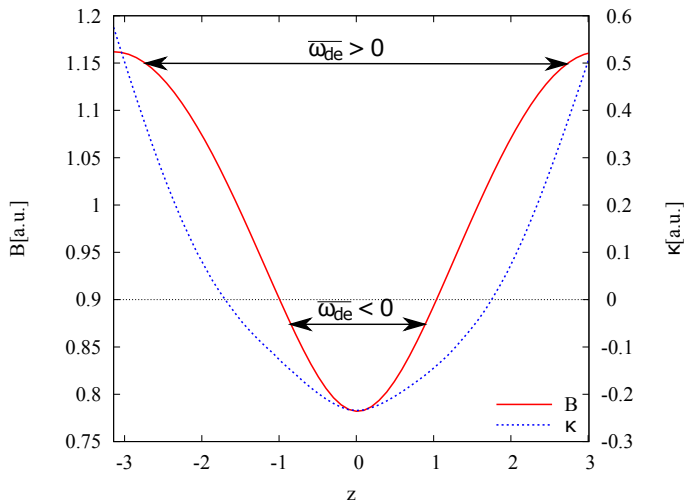}
\caption{\label{fig:badcurvature} The magnetic field strength $B$ (left axis) and the local curvature $\kappa$ (right axis) along a field line of the half-flux flux surface of the DIII-D tokamak. The two arrows indicate the paths of two trapped electrons with
different pitch angle $\lambda$ and thus with different bounce points. The deeply trapped particle only samples bad (i.e. negative) local curvature along its path, therefore its bounce averaged curvature will also be negative. 
A barely trapped particle that samples both good and bad curvature might have good bounce averaged curvature. More particles are trapped in regions of bad local curvature if the two curves of magnetic field and local curvature are in phase.}
\end{figure}

The stability analysis of TEMs in general geometry via a dispersion relation is not as accessible as in tokamaks. However, it is possible to define a rate of gyrokinetic energy transfer $P_e$ from the fluctuating electric field to the electrons \cite{Proll2012, Helander2012, Helander2013}, which, at the point of marginal stability where the growth rate $\gamma$ approaches zero, can be written as
\begin{equation}
 P_e = \frac{\pi e^2}{T_e} \int \frac{\mathrm{d}l}{B}\int \mathrm{d}^3v \delta(\omega - \overline{\omega}_{de}) \overline{\omega}_{de} (\overline{\omega}_{de} - \omega_{\ast e}^T)|\overline{J_0 \phi}|^2 f_{e0}.
 \label{eq:Pe}
\end{equation}
Here, $T_e$ is the electron temperature, $\omega$ is the real frequency of the mode and $\overline{\omega}_{de}=\overline{\textbf{k}_{\bot}\cdot \textbf{v}_{de}}$ denotes the bounce-averaged precessional drift frequency of the electrons, whose drift velocity is given by 
$\textbf{v}_{de}$. The velocity-dependent diamagnetic frequency is given by $\omega_{\ast e}^T=\omega_{*e}\left[1+\eta_e(\frac{\mathcal{E}}{T_e}-\frac{3}{2})\right]$, where $\mathcal{E}$ denotes the energy of the particle, $\eta_e=\left.\frac{\mathrm{d}\ln T_e}{\mathrm{d}r}\right/\frac{\mathrm{d}\ln n}{\mathrm{d}r}$ gives the ratio between the scale lengths of electron temperature gradient and density gradient and the diamagnetic frequency is defined as $\omega_{*e}=\frac{T_e}{neB}\left(\textbf{B}\times \textbf{k}_{\bot}\right)\cdot \nabla n$. 
In addition, $J_0$ denotes the Bessel function, $\phi$ is the electrostatic potential and $f_{e0}$ is the Maxwellian distribution function of the electrons.
For the electrons to have a destabilising influence, the energy transfer rate must be negative, $P_e<0$. This means that $\overline{\omega}_{de}\omega_{*e}>0$ (if the temperature gradient is small, $\eta_e<2/3$) at least for some particles in velocity space, because all the other terms are positive definite. Thus, for some of the trapped electrons, the precessional drift must be resonant with the propagation of drift waves. In which direction the trapped electrons precess depends on the curvature they sample along their path in the magnetic field: 
\begin{equation}
\overline{\omega_{de} }(\lambda)\propto \int _{z_1}^{z_2}\frac{ \kappa \left(1-\lambda B(z)/2\right) }{\sqrt{1-\lambda B(z)}} \mathrm{d}z,
\end{equation}
where the integration is taken along the particle path along a field line, with the pitch angle like coordinate $\lambda = v_{\bot}^2/v^2B$, bounce points $z_i$ and, most importantly, the local radial curvature $\kappa$, which can have positive and negative values, depending on the direction of the drift.
Here, the diamagnetic frequency is chosen to be negative, $\omega_{*e} <0$, which means that for a resonance to exist the precessional drift must also be negative, $\overline{\omega_{de} } (\lambda)<0$. 
For a particle to assume such a negative precessional drift it must sample mainly negative local curvature along its path, see Fig. \ref{fig:badcurvature} - so-called ``bad curvature''. This local ``bad curvature'' has long been recognised as the drive for interchange instabilities \cite{Jenko2001a, Kadomtsev1966} and ITGs, and it does play an important role for TEMs, but there its average over the bounce motion determines the stability properties.
Many of the trapped particles will have averaged bad curvature $\overline{\omega_{de} } (\lambda)<0$ if the particles are mainly trapped in regions of local bad curvature, i.e. if the magnetic field and the local curvature are in phase. Magnetic
configurations where this is the case should therefore be characterised by destabilising electrons, $P_e<0$, and should thus be prone to TEM instabilities. 
Configurations where the magnetic field and the local curvature are even partially out of phase, on the other hand, should have reduced TEM activity. (Another possibility to achieve mainly good average curvature would be to improve the local curvature alltogether, for example by having a high plasma pressure $\beta$ \cite{Bourdelle2003}, but our goal is to also optimise the vacuum configurations, so increasing $\beta$ is not an option.)
In the limit where \emph{all} particles experience good average curvature, $\overline{\omega_{de} } (\lambda)>0$, as is the case in
quasi-isodynamic stellarators \cite{Gori1996, Subbotin2006} with the maximum-$J$-property ($J$ is the action integral of the bounce motion of trapped particles and constant on flux surfaces, the maximum of $J$ being at the plasma centre), it can be shown that TEMs and trapped-particle modes
are stable in large regions of parameter space, i.e. if the electron temperature gradient is small, $\eta_e<2/3$. 
\subsection{Linear simulation results}
\begin{figure*}
\includegraphics[width=1.0 \textwidth]{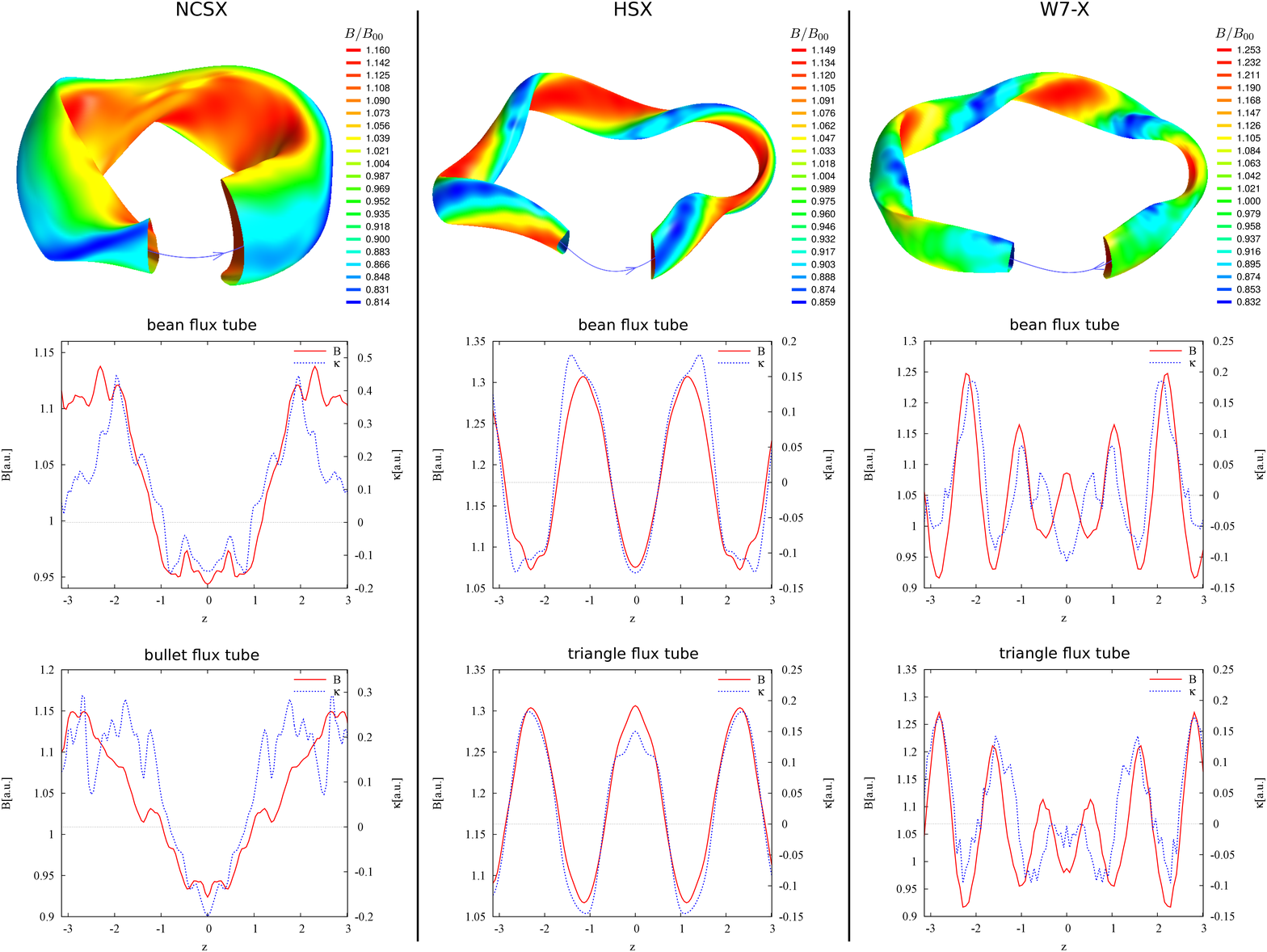}
\caption{\label{fig:thegeometries}A comparison of the geometries of the three simulated stellarators NCSX, HSX and W7-X. Displayed are the magnetic field strength at the outermost flux surface at the top and the magnetic field strength and local curvature (on the left and right axis, respectively) along the two simulated flux tubes per configuration, each of them at half flux $s=0.5$.}
\end{figure*}

\begin{figure}
\includegraphics[width= 0.3 \textwidth, angle=270]{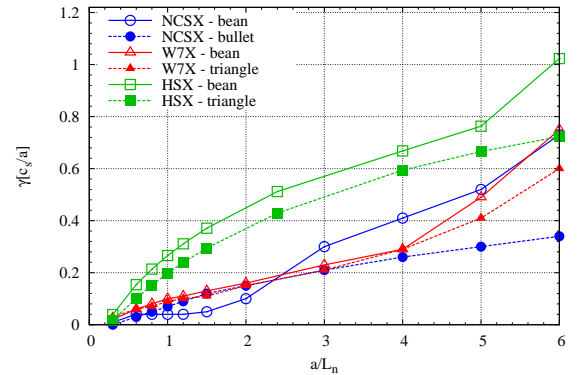}
\caption{\label{fig:linearresults}Linear growth rates of density-gradient-driven TEMs in each of the simulated flux tubes in NCSX, HSX and W7-X. At each simulated density gradient $a/L_n$, where $a$ is the minor radius of the device and $L_n$ the density gradient scale length, the growth rate of the most unstable mode is displayed.}
\end{figure}

Configurations with most of the particles experiencing good average curvature can also benefit from enhanced TEM stability, as can be shown with linear simulations. The simulations are performed with the GENE code \cite{Jenko2000} in the collisionless and electrostatic limit.
The geometry of the different configurations is incorporated into GENE via the GIST geometry interface \cite{Xanthopoulos2009}, and we chose to study three very different stellarator equilibria: the quasi-axisymmetric stellarator design NCSX (National Compact Stellarator Experiment, nowadays designated QUASAR \cite{Neilson2014}, see Fig.~\ref{fig:thegeometries} on the left), the quasi-helically symmetric stellarator experiment in Madison, Wisconsin, HSX (Helically Symmetric Experiment, see Fig.~\ref{fig:thegeometries} in the middle), 
and the stellarator Wendelstein 7-X (W7-X, see Fig.~\ref{fig:thegeometries} on the right), which approaches quasi-isodynamicity. For each of the configurations two stellarator-symmetric flux tubes were chosen from the flux surface at half toroidal flux, $s=0.5$, - one where the binormal coordinate $\alpha=0$ in the midplane, and the second one at $\alpha=\pi/N$ where $N$ denotes the number of periods. 
In all configurations, the flux tube with $\alpha=0$ is centered around the bean-shaped poloidal cross section and is therefore referred to as ``bean flux tube''. The poloidal cross section at the center of the $\alpha=\pi/N$ flux tube is either triangle-shaped (in W7-X and HSX) or bullet-shaped (in NCSX) and therefore called ``triangle flux tube'' or ``bullet flux tube'', respectively.
The two quasi-symmetric devices NCSX and HSX display a strong overlap of the magnetic trapping well and the region of bad local curvature. This is not the case though for W7-X, especially at the centre of the flux tube. The analytical theory therefore suggests that W7-X should have lower TEM growth rates than both NCSX and HSX.
On the other hand, it should be noted that in NCSX, the region of bad local curvature, though it overlaps with the magnetic trapping well, is very small compared with the large region of good local curvature, especially in the bullet flux tube. NCSX might therefore
benefit from enhanced TEM stability, too.
We simulated purely density-gradient-driven TEMs, thus choosing both ion and electron temperature profiles to be flat. For each value of the normalised density gradient $a/L_n$, where $a$ denotes the minor radius and $L_n^{-1}=-\frac{\mathrm{d}\ln n_a}{\mathrm{d}r}$ the density gradient scale length, several wave numbers $k_y\rho_s$ ($\rho_s$ is the ion sound radius) around the expected most unstable mode were simulated, and the highest growth rate was then recorded. 
The predicted behaviour of the different configurations is indeed born out in the simulations: W7-X and NCSX have the lowest linear growth rates, whereas HSX has the highest, see Fig.~\ref{fig:linearresults}. The bullet flux tube of NCSX is more stable than the bean flux tube, which can be explained by the region of bad local
curvature being even smaller in the bullet flux tube than in the bean flux tube. The fact that the bean flux tube in HSX has higher growth rates than the triangle flux tube can be attributed to the fact that there is a magnetic trapping region with bad local curvature at zero ballooning angle, which should enhance the mode. This is very much in line with the analytical predictions and previous linear simulation results \cite{Proll2013}.
To summarise this section: it was expected from analytical calculations and also shown via gyrokinetic simulations that configurations where fewer particles have average bad curvature benefit from enhanced stability of density-gradient-driven TEMs. Very recent nonlinear results 
confirm these findings \cite{Helander2015}. More extensive nonlinear data will be published in a later paper. Based on this knowledge about the TEM it should be possible to come up with a measure, a so-called proxy, that only depends on the geometric properties of a configuration and that will give an estimate 
of the linear (and ultimately also nonlinear) TEM activity of the respective configuration.

\section{Proxy function}
\label{sec:proxies}
\begin{figure}
\includegraphics[width= 0.5 \textwidth]{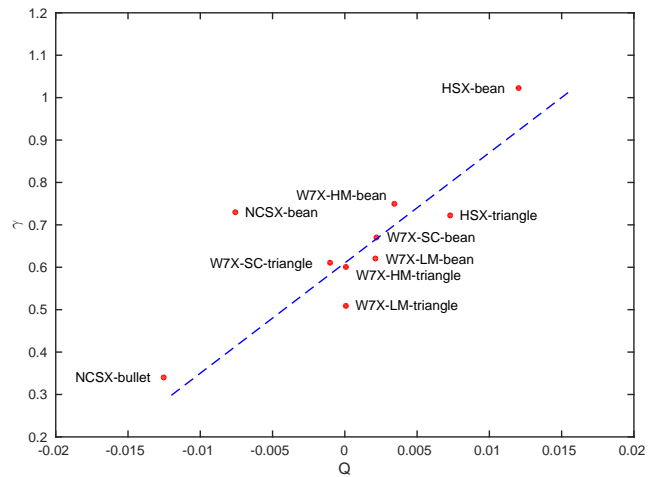}
\caption{\label{fig:Qadvanced}The comparison between the value of the proxy, $Q_{\rm bounce}$, here displayed as $Q$, and linear growth rate $\gamma$ at a density gradient $a/L_n=6$, where the maximum over various binormal wave numbers $k_y$ was found. 
Shown are the values for all of the flux tubes discussed above and additional flux tubes for the low-mirror (LM) and standard configuration (SC) of W7-X, where the high mirror (HM) is the one that has been used throughout this paper. 
There is a clear correlation between proxy value and linear growth rate between the different optimisation families. Note: the high density gradient was chosen to ensure the appearance of a strong TEM, especially for the fairly stable configurations like W7-X. 
As can be seen in Fig.~\ref{fig:linearresults}, the relative stability of different configurations remains the same for the higher gradients.}
\end{figure}

The aim of the proxy function is to provide a means of estimating the stability of a configuration towards TEMs \emph{efficiently}, so that the calculation can be performed for many different equilibria during the process of the optimisation. 
We remember the analytical expression for the gyrokinetic energy transfer rate (Eq.~(\ref{eq:Pe})) and use this as inspiration for our proxy function.
The central finding from the analytical theory discussed above was that it is beneficial for a configuration to have as few trapped particles as possible with bad average curvature. 
In order to obtain the improved proxy function $Q_{\rm bounce}$ we thus average the bounce averaged curvature $\overline{\kappa} \propto \overline{\omega}_d(\lambda)$ of a particle with pitch angle $\lambda$ over all trapped particles, i.e. over all pitch angles $\lambda$, equivalent to how the average is done in Eq.~(\ref{eq:Pe}):
\begin{equation}
 Q_{\rm bounce }= -\int_{1/B_{\mathrm max}}^{1/B_{\mathrm min}}\overline{\omega}_d(\lambda)\mathrm{d}\lambda,
\end{equation}
with 
$$\overline{\omega}_d(\lambda)=\int_{-\ell_0}^{+\ell_0}H\left(\frac{1}{\lambda}-B(\ell)\right)\omega_d(\lambda, \ell)\frac{\mathrm{d}\ell}{\sqrt{1-\lambda B(\ell)}}$$ and
 where the $B_{\mathrm min}$ and $B_{\mathrm max}$ denote the minimum and maximum of the magnetic field along a given field line.
The minus sign is introduced to make a configuration with a majority of particles with good average curvature have a minimum $Q_{\rm bounce}$, which seems more intuitive from an optimisation point of view. If we plot the maximum growth rate obtained
from TEM simulations with a pure density gradient for various configurations and flux tubes versus the corresponding proxy value we see that the proxy correlates well with the linear growth rates, see Fig.~\ref{fig:Qadvanced}.
Especially if two configurations are very different, the proxy correctly predicts which one is the more stable.
For configurations that are very similar, however, for example different W7-X configurations that mainly differ by their mirror ratio (HM being high mirror, LM being low mirror, and SC being the standard configuration), a lower proxy value does not necessarily mean a lower TEM growth rate.
This means an optimisation will probably need to make large steps in proxy value $Q$ to ensure that the found optimised equilibrium indeed has lower levels of TEM activity.
\section{The proof-of-principle configuration}
\label{sec:HSXopt}
\begin{figure*}
\includegraphics[width=\textwidth]{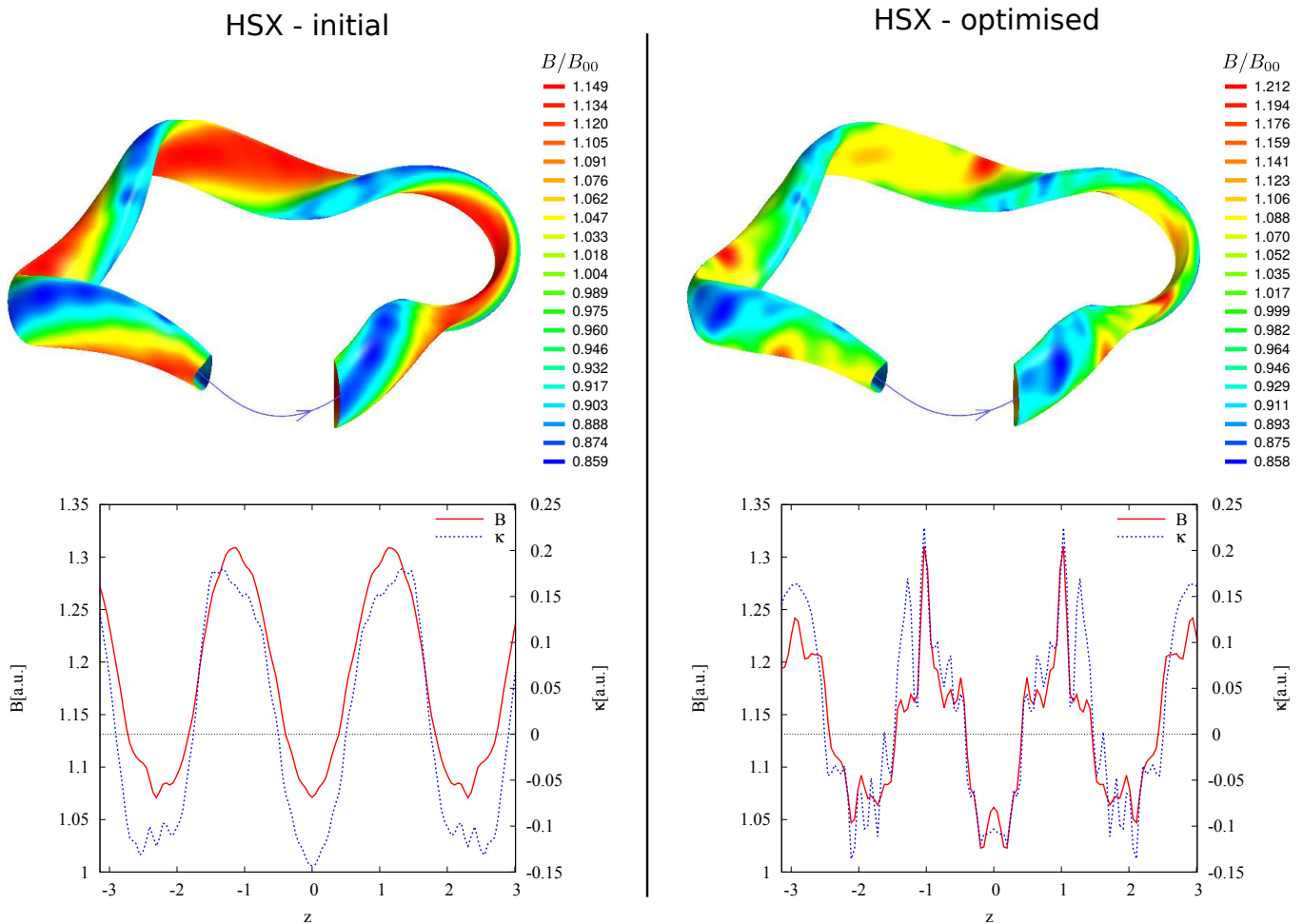}
\caption{\label{fig:HSXoptinitcomp}Comparison of the initial helically-symmetric HSX and the derived optimised equilibrium produced with STELLOPT. Shown are the magnetic field strength on the outermost flux surface at the top and the magnetic field strength and local bad curvature (on the left and right axis, respectively) along the bean flux tube at the surface with half flux $s=0.5$. The optimised equilibrium is not helically symmetric anymore.}
\end{figure*}

\begin{figure*}
\includegraphics[width=0.5 \textwidth, angle=270]{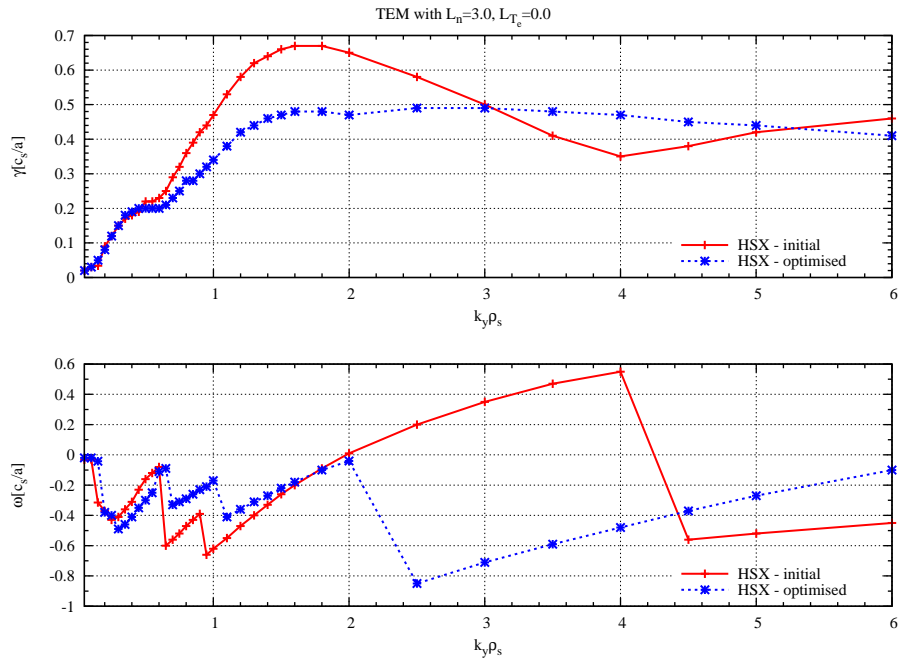}
\caption{\label{fig:optikyscan}A comparison of the linear TEM growth rates $\gamma$ and the real frequencies $\omega$ in the bean flux tubes of the initial HSX equilibrium and the optimised equilibrium for different wave numbers $k_y\rho_s$ at a fixed density gradient $a/L_n=3$. The optimised equilibrium has lower growth rates at the turbulence relevant scales.}
\end{figure*}

\begin{figure}
\includegraphics[width=0.3 \textwidth, angle=270]{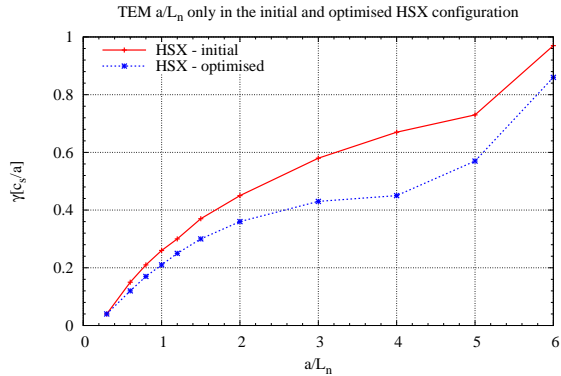}
\caption{\label{fig:optigammamax}A comparison of the linear TEM growth rates in the bean flux tubes of the initial HSX equilibrium and the optimised equilibrium at different density gradients $a/L_n$. The optimised equilibrium has reduced growth rates for all density gradients.}
\end{figure}

A first attempt at an optimisation was made with HSX as the starting equilibrium. For this first proof-of-principle optimisation STELLOPT's fixed boundary mode was chosen, which means the accessible configuration space was very large. 
This was indeed necessary.
\begin{table}
\begin{tabular}{|l|l|l|l|}
\hline
Optimised quantity    & Target Count & Target & Weight $1/\sigma$ \\
\hline
Neoclassical Transport $\epsilon_{\rm eff}^{3/2}$ & 127 & 0 & $0.001$ \\
Turbulent Transport $Q_{\rm bounce}$& 25 & 0 & $1000$ \\
Major Radius $R_0$& 1 & 1.22 & $10$ \\
\hline
\end{tabular}
\caption{Targets for HSX optimisations.  The turbulent and neoclassical values are evaluated at multiple radial locations.}
\label{tab:HSXopt1122}
\end{table}
The constraints of fixed aspect ratio and low neoclassical transport prevented STELLOPT from finding an equilibrium with lower proxy, i.e. better average curvature. In order to see any change in the proxy the requirement of low neoclassical transport had to be relaxed significantly, which means
that the weight for the neoclassical transport was chosen to be very small compared with the weight for our proxy (see Table \ref{tab:HSXopt1122}). The resulting TEM-optimised equilibrium shown in Fig.~\ref{fig:HSXoptinitcomp} on the right has lost the helical symmetry.
This leads to a significant increase in the neoclassical transport - the neoclassical effective ripple went up by an order of magnitude. Moreover, the magnetic field along the field line of this preliminary equilibrium is very jagged. 
This should of course be avoided when trying to find a truly optimised configuration. In this case, however, our primary focus is to show that the proxy works. To test this, we first performed linear GENE simulations. 
A scan over the binormal wave vector $k_y\rho_s$ for a purely density-gradient-driven TEM with a density gradient $a/L_n=3$ and no temperature gradient shows that the linear growth rates are indeed reduced for the optimised equilibrium, 
at least for the scales where turbulence is generated (Fig.~\ref{fig:optikyscan}).
This stabilisation also holds for a large range of density gradients, as can be seen in Fig.~\ref{fig:optigammamax}. 
For these simulations, a scan over various wave numbers was performed and the highest growth rate for each gradient is displayed. These linear results lead to the expectation that a nonlinear simulation of this proxy-optimised configuration
would also show reduced transport. 
A test with pure density-gradient-driven TEM turbulence at a density gradient of $a/L_n=3$ shows that the nonlinear electron heat flux went from $Q/Q_{GB}= 1.05$ to $Q/Q_{GB}=0.62$, where the heat fluxes are measured in Gyro-Bohm units $Q_{GB}=nT_ec_s\rho_s^2/a^2$, with the density $n$, the ion sound speed $c_s$, and the sound Larmor radius $\rho_s$. 
This means a reduction of about $40\%$ was achieved. 
\begin{figure}
\includegraphics[width=0.5\textwidth]{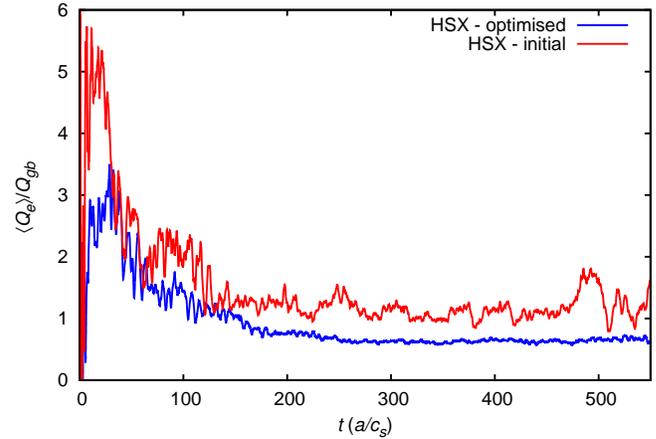}
\caption{\label{fig:optiNL}A comparison of the nonlinear heat flux in the bean flux tubes of the initial HSX equilibrium and the optimised equilibrium at a density gradient $a/L_n=3$. The grid for these simulations was chosen as follows: 
in the radial direction $nkx=192$, in the binormal direction $nky=48$ with the lowest wave number being $kymin=0.01$, along the field line $nz0=64$ for the initial configuration or $nz0=256$ 
for the optimised configuration, respectively, for the parallel velocity $nv0=40$ and for the magnetic moment $nw0=20$.}
\end{figure}
However, the high neoclassical transport remains a handicap, and, since the fixed boundary mode of STELLOPT was chosen to create this optimised equilibrium, the magnetic field is not realisable by simply adjusting the currents in the existing coils of the HSX experiment. 
Additional STELLOPT runs should therefore be used to try to find actually optimised but experimentally realisable configurations using the free-boundary mode. 
\section{Conclusions and outlook}
\label{sec:conclusions}
In this paper we have presented a method to optimise stellarators for density-gradient-driven TEM turbulence using the optimisation code STELLOPT. The optimisation for TEM turbulence complements ongoing efforts to optimise stellarators not only for neoclassical transport but also for turbulent transport. 
We used analytical theory and linear flux-tube simulations performed with the GENE code to guide us in devising a proxy function that can stand in for the expected turbulent heat flux. Both analytical theory and the linear simulations suggested
that configurations with a lower fraction of particles with bounce-averaged bad curvature should be less unstable to TEMs.
The bounce-averaged curvature averaged over all trapped particles was thus chosen for the proxy. The comparison between the proxy and linear growth rates for TEMs in various configurations revealed that the proxy is well suited to predict the relative stability of a configuration.
Assuming that the linear growth rates are correlated with the turbulent transport levels, the proxy should thus be able to guide the optimiser towards configurations with lower TEM turbulence levels.
A first proof-of-principle configuration where this was indeed achieved was presented. There, the linear growth rates were reduced compared with the starting equilibrium of HSX, as was the turbulent heat. This configuration was, however, not realisable
with HSX's given coil set. The presentation of an experimentally feasible TEM-optimised configuration is deferred to a future publication.\\
One possible improvement of the current proxy would be to include further weighting of the deeply trapped particles by taking into accound the mode structure of the linear modes via the electrostatic potential $|\phi|^2$, as it is included also in the equation for the energy transfer rate, Eq.~(\ref{eq:Pe}).
Another possibility that has proven fruitful in the reduction of ITG turbulence is to include the distance of the flux surfaces in the optimisation, trying to find configurations where this distance is particularly large, which would result in a smaller effective gradient
and thus possibly large regions of stability in parameter space. Combining the TEM optimisation with the ITG optimisation where the local curvature is minimised might also lead to better results for TEM turbulence. 
The simultaneous reduction of both ITGs (and interchange instabilities in general) and TEMs might already be happening with our proxy, when the reduction of the bounce averaged bad curvature is achieved by making the local curvature better. 
However, there might be configurations where this is not feasible, but shifting the local bad curvature away from the magnetic wells is.\\
In addition to improving the proxy function and thus hopefully reducing the turbulence at high gradients other problems in turbulence optimisation could and should be addressed. 
One of these problems is to increase the critical gradient for the onset of turbulence, which is particularly important if the turbulence is very ``stiff'', i.e. if the heat flux increases dramatically once the critical gradient is exceeded.
Ideally, in addition to reducing the turbulent heat flux, one would achieve an increase in the particle flux to flush out impurities from the plasma, but a deeper understanding of turbulence is required before this challenge can be tackled. 
In general it remains to be seen to what extent the simultaneous optimisation of several different aspects (neoclassical and turbulent transport, fast-particle confinement, divertor etc.) can be successful.
Until then the TEM optimisation presented here will serve as a useful tool to learn more about the influence of geometry on TEM stability and will certainly guide future efforts.
\begin{acknowledgments}
The authors thank T. G\"orler, G.W. Hammett, P. Helander, D.R. Mikkelsen, J.N. Talmadge and M.C. Zarnstorff for many fruitful discussions as well as Yu. Turkin for providing the MCviewer for displaying the magnetic geometry and S.P. Hirshman for access to the VMEC code.\\
Some of these simulations were performed on the HELIOS supercomputer, Japan. One of the authors (J.H.E. Proll) gratefully acknowledges funding from the Max Planck/Princeton Center for Plasma Physics. This work has been carried out within the framework of the EUROfusion
Consortium and has received funding from the Euratom research and training programme 2014-2018 under the grant agreement No 633053. The views and opinions expressed herein do not necessarily reflect those of the European Commission.
\end{acknowledgments}

\end{document}